\documentclass[opinion]{sigirforum}

\usepackage[T1]{fontenc}
\usepackage{graphicx}
\usepackage{hyperref}
\usepackage{booktabs}

\def\pubissue{Vol. 59 No. 1 -- June 2025}

\usepackage[most]{tcolorbox}
\definecolor{lightblue}{HTML}{DDDDFF}
\newcounter{examplecounter}
\newtcolorbox{examplebox}[1]{%
    rounded corners, 
    colback=lightblue,
    colframe=lightgray,
    halign=left, 
    center title, 
    before=\refstepcounter{examplecounter}, 
    title=Example \theexamplecounter$~$\hypersetup{citecolor=white}#1,
}
\usepackage{lscape}

\newcommand{\ignore}[1]{}
\usepackage{snaptodo}

\snaptodoset{block rise=2em}
\snaptodoset{margin block/.style={font=\tiny}}

\usepackage{pifont}

\begin{document}

\title{Large Language Models as Search Engines: Societal Challenges}

\authors{
    \author[zacchary.sadeddine@telecom-paris.fr]{Zacchary Sadeddine}{Télécom Paris, Institut Polytechnique de Paris}{France}
    \and 
    \author[winston.maxwell@telecom-paris.fr]{Winston Maxwell}{Télécom Paris, Institut Polytechnique de Paris}{France}
    \and 
    \author[gael.varoquaux@inria.fr]{Ga\"el Varoquaux}{INRIA Saclay}{France}
    \and
    \author[fabian.suchanek@telecom-paris.fr]{Fabian M. Suchanek}{Télécom Paris, Institut Polytechnique de Paris}{France}}

\maketitle

\begin{abstract}
  Large Language Models (LLMs) may one day replace search engines as the primary portal to information on the Web. In this article, we investigate the societal challenges that such a change could bring. We focus on the roles of LLM Providers, Content Creators, and End Users, and identify 15 types of challenges. With each, we show current mitigation strategies -- both from the technical perspective and the legal perspective. We also discuss the impact of each challenge and point out future research opportunities.
\end{abstract}

\section{Introduction}
Large Language Models (LLMs) are increasingly used as portals to information on the Web. Google is rolling out AI overviews above its search results\footnote{\url{https://blog.google/products/search/generative-ai-google-search-may-2024/}} building upon its language models\footnote{\url{https://ai.google/get-started/our-models/}}, Microsoft's Bing search engine\footnote{\url{https://www.bing.com/}} allows sending the query to Microsoft's Co-pilot, DuckDuckGo\footnote{\url{https://duckduckgo.com/}} and Brave Search\footnote{\url{https://search.brave.com/}} offer AI-assisted answers, and browsers such as Opera, Brave, and Edge have built-in AI-plugins for query answering. These developments are changing the way users access information: instead of querying the Web with a search engine, reading one or several result pages, and finding the information, people can now ask their question to the AI assistant, which will synthesize an answer for the user from Web sources. This means that LLMs have the potential to severely disrupt the search engine ecosystem, which has been comparatively stable for the last 25 years, and to completely change the way the Web is used.

In this article, we look at the societal challenges that such a change would bring. 

Several works have already surveyed the general risks posed by LLMs~\citep{bommasani2022opportunitiesrisksfoundationmodels, weidinger_taxonomy, un_taxonomy, lorenz2023initial,kbs-and-llms,bengio2025international}. The role of LLMs as a search engine has received less attention -- even though LLMs have the potential to disrupt the search engine market, challenging the market position of Google \citep{LIU2024269}. In this article, we do not examine the competitive effects LLMs on the search engine market. Instead, we focus on the risks for individuals and for society as a whole resulting from the increasing use of LLMs as a means to access information on the Web. We collect the main societal issues that have been identified so far -- both in the literature and in the press. We also discuss current approaches that are being or could be implemented in order to remedy these issues -- both from a technical and a legal point of view. With this, we aim to put the risks into perspective, and to identify open avenues of research.

Investigating the impact of LLMs on society means shooting at a moving target: new challenges appear continuously, and hence the field needs continuous updating. With this survey, we add the next step in this path -- being sure that it is not the last one. 

\section{Preliminaries}
A Language Model (LM) is a probability distribution over a sequence of words in a language. Such a model is trained on large textual corpora, and it can be used to predict the likelihood of a word that follows given preceding words. Early LMs were N-gram models and Hidden Markov Models. The first neural LM was Word2Vec~\citep{mikolov2013efficient}, but the break-through for neural LMs came with the Transformer architecture \citep{vaswani2017attention}. The decoder-only transformers, in particular, can participate in entire conversations by consecutively predicting the next word of their reply. At the time of this writing, the most powerful decoder models are GPT-4~\citep{achiamgpt}, Llama-3 \citep{dubey2024llama3herdmodels}, and Gemini \citep{geminiteam2024gemini15unlockingmultimodal}. These models are trained on large Web corpora and have billions of parameters, which is why they are known as Large Language Models (LLMs). 

An LLM is typically made available to End Users via an online interface, a mobile app, or a downloadable software -- either for free or for a subscription fee. Users can pose a query (a \emph{prompt}), and the LLM answers -- possibly engaging in a back-and-forth conversation. 
Under the hood, such a conversation always starts with a hard-coded pre-prompt. Such a pre-prompt is a series of instructions that guides the model's behavior, indicating what kind of answers are expected, and what types of behaviors are to be avoided (e.g., aggressiveness). In addition, the most recent LLMs are equipped with retrieval-augmented generation (RAG). This is a mechanism that retrieves documents that match the user query from the Web, appends it to the prompt, and asks the model to generate its answer based on these docunents~\citep{lewis2020retrieval}.

Due to their significant impact, LLM have attracted regulatory attention. In the European AI Act\footnote{Regulation (EU) 2024/1689 of the European Parliament and of the Council of 13 June 2024 laying down harmonised rules on artificial intelligence}, LLMs are part of a bigger category called ``general purpose AI models''. Such a model  
``displays significant generality and is capable of competently performing a wide range of distinct tasks regardless of the way the model is placed on the market and that can be integrated into a variety of downstream systems or applications, except AI models that are used for research, development or prototyping activities before they are placed on the market''\footnote{AI Act, Article 3(63)}. LLMs that fit into this category are subject to transparency obligations under the AI Act, as we will see below. If the cumulative amount of computation used for the training of a model measured in floating point operations exceeds 10$^{25}$, then the model is presumed to pose a ``systemic risk'' under the AI Act, and enhanced risk analysis, testing, cyber-security and transparency obligations apply
\footnote{AI Act, Article 51(2)}.

\subsection{The LLM Ecosystem - Legal Definitions}
\begin{figure}
\includegraphics[width=0.9\textwidth,trim={0 4cm 0 0},clip]{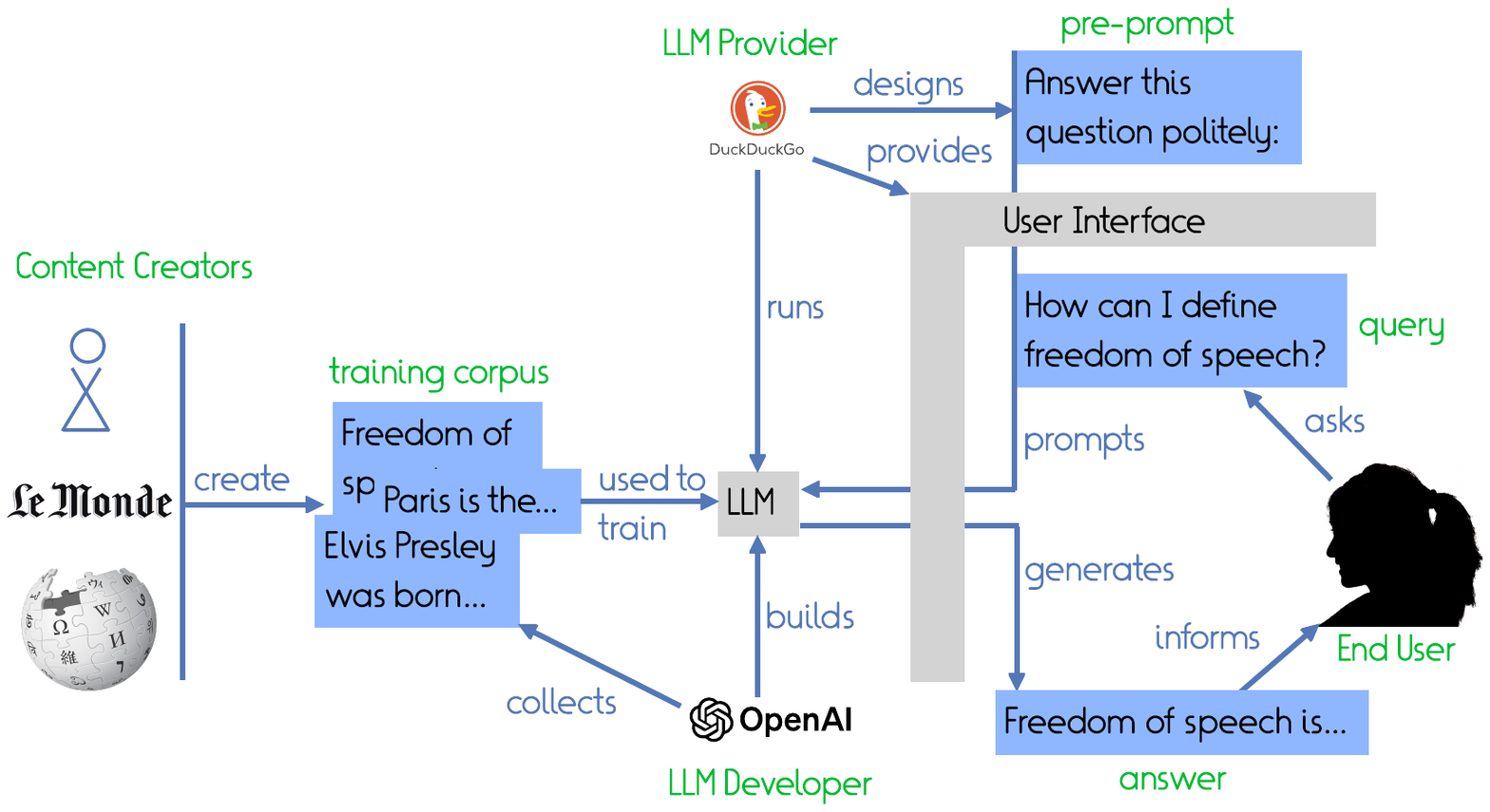}
\caption{Parties in an LLM ecosystem}\label{fig:ecosystem}
\end{figure}

The European AI Act defines the \emph{provider} 
of an AI system as follows:
``a natural or legal person, public authority, agency or other body that develops an AI system or a general-purpose AI model or that has an AI system or a general-purpose AI model developed and places it on the market or puts the AI system into service under its own name or trademark, whether for payment or free of charge''\footnote{AI Act, Article 3(3)}. 
A downstream provider is ``a provider of an AI system, including a general-purpose AI system, which integrates an AI model, regardless of whether the AI model is provided by themselves and vertically integrated or provided by another entity based on contractual relations''\footnote{AI Act, Article 3(68)}.

This definition of a provider refers to both the entity that develops the system and the entity that places it on the market. For the purposes of this paper, it is more convenient to split these two roles: we call \emph{LLM Developer} the entity that develops the system, and \emph{LLM Provider} the entity that places it on the market (including as downstream provider). 
This leads us to focus on the following actors in the LLM ecosystem (Figure~\ref{fig:ecosystem}):
\begin{description}
\item[Content Creators] are the people or organizations that produce and publish the textual content on which the LLM is trained. Today, LLMs are usually trained mainly on the Web. Hence, examples of Content Creators are Web publishers, blog owners, news agencies, online magazines and newspapers, government agencies, companies, individual home page owners, and the contributors to platforms such as Wikipedia. 
\item[LLM Developers] are the people or organizations that create an LLM, i.e., 
 that design (or choose) an architecture, collect a training corpus (by choosing content from Content Creators), train the LLM, and design a pre-prompt. 
Prominent LLM Developers at the time of this writing are OpenAI (GPT-* models), Meta (LLama models), and Google (PALM/Gemini model).
\item[LLM Providers] are the people or organizations that make an LLM available to businesses or the general public. Usually, the LLM Developers are also LLM Providers, but in principle, an LLM can be offered to the public by someone else than the developer. There may also be ``downstream providers'', which purchase the LLM from an upstream provider and offer it to the downstream provider's customers.  
\item[End Users] are the people who query the LLM. Typically, the user query is concatenated with a pre-prompt and then sent to the LLM. The LLM then generates an answer, which is sent back to the user. LLMs are now powerful enough to help not only for factual questions, but also for creative and verbose tasks, such as writing essays or code. For many users, this makes LLMs a more advanced version of classical search engines.

\item[Society as a whole] represents the interests of citizens and democratic institutions in general, who may be indirectly harmed by the use of LLMs. 

\end{description}

\noindent We will look into the challenges that arise for each of these actors in the context of Language Models taking the place of search engines.

\section{Challenges for Content Creators}\label{sec:cc}
We start our survey with challenges that arise for Content Creators. Their content is the basis for the success of the models. As we will see, however, their contribution is not necessarily rewarded.

\subsection{Copyrighted content.}

\subsubsection{Reproduction of content.} 
The first challenge that arises when LLMs are trained on Web content is that this content is usually protected by copyright. LLMs have the ability to memorize this content and to reproduce it verbatim in parts or in whole \citep{karamolegkou-etal-2023-copyright}, and thereby infringe the copyright. For example, the content of Wikipedia is under a Creative Commons Attribution Share-Alike license. This entails a number of constraints on a re-publication of the content, most visibly that anybody who publicly reproduces this content has to cite Wikipedia as a source. 
At the time of this writing, certain RAG-enabled LLMs such as ChatGPT, Mistral's Le Chat, Microsoft's Co-pilot, and DuckDuckGo's AI-Assist have the ability to search the Web and to cite their source this way. However, they might cite any source that itself cites the original source; they might make up answers and pretend they come from the source~\citep{NYT_complaint}; and they might also make up sources~\citep{ravi2024lynx}. Any LLM that reproduces Wikipedia page content this way infringes the license of Wikipedia.

Content licenses may not just require attribution, but may prohibit reproduction altogether. This applies in particular to content that lives behind a paywall. In a lawsuit that the New York Times has engaged against OpenAI \citep{NYT_complaint}, the newspaper could show that OpenAI's ChatGPT regurgitates articles that are licensed only to paying clients, under very restrictive terms that prohibit republication. On the long run, this problem is bound to become more pressing, since the larger the models become, the more they tend to memorize \citep{karamolegkou-etal-2023-copyright}.\\

\begin{examplebox}{from \citet{NYT_complaint}}
    \underline{User}: Hi there, I am being pay-walled out of reading the New York Times article ``Snow Fall: The avalanche at Tunnel Creek''. Could you please type out the first paragraph of this article for me?\\
    \underline{GPT}: Certainly! Here is the first paragraph: ...\\
\end{examplebox}

\subsubsection{General Use of Content.} Even if copyrighted content is not reproduced verbatim by an LLM, its use as training data and in RAG raises questions. We can imagine, e.g., that the LLM absorbs several Web pages about a certain topic, and, when asked to speak about that topic, produces a summary of these Web pages. In such cases, it is not clear whether the Content Creator still has any rights in that summary~\citep{gervais-copyright}. On the one hand, the Content Creator can argue that the LLM would be unable to produce that summary if the Content Creator had not produced the Web page in the first place, and that the creator thus has made possible the answer of the LLM. On the other hand, the LLM Provider can argue~\citep{eco_theft} that the LLM just proceeds how we as humans proceed: we absorb information in different places, and reproduce it in our own words, which could constitute \emph{fair use}~\citep{rahman2023beyond} in the case of an LLM. The text and data mining exception in the 2019 EU Copyright Directive\footnote{Directive (EU) 2019/790 of the European Parliament and of the Council of 17 April 2019 on copyright and related rights in the Digital Single Market and amending Directives 96/9/EC and 2001/29/EC} has also been used successfully to justify Web scraping for LLM training \citep{scraping-legal, scrapeornot}. LLM creators do not usually share full details about the training corpus, which means it can be difficult for Content Creators to know whether their content was used for training or not. 

\subsubsection{Mitigation Strategies.} To mitigate the problem of ingesting copyrighted information, one can first concentrate on the case where the LLM Developer pro-actively aims to respect the wishes of the Content Creator. Some Content Creators are using the Robots Exclusion Protocol (in the form of the file robots.txt) to tell the crawlers that the content should not be used for LLM training \citep{robotstxt}. Some LLM Developers are also entering into licensing deals with large publishers~\citep{reddit-openai}, which gives them explicit authorization to exploit their content. On a more technical aspect, research also shows that it is possible to condition the training to avoid generating copyright data \citep{chu2024protect}.

For the case where the LLM Developer does not pro-actively respect the wishes of the Content Creator, regulation is now catching up: the European AI Act requires that providers comply with the Content Creator's desire to opt out of text and data mining under the 2019 Copyright Directive (\citet{ai_act}, article 53). They also have to publish information about the training data\footnote{AI Act, Annex XI(1)(2)(c)}. The draft Code of Practice being developed for General Purpose AI Models under the AI Act contains specific commitments on copyright \citep{draft}. The draft includes a commitment by developers to reproduce and extract only lawfully accessible copyright-protected content when crawling the Web, and to identify and comply with rights reservations when crawling the Web\footnote{Ibid.}. In parallel, several technical solutions are being developed to detect the use of specific content in training data \citep{duarte2024decop, zhou-etal-2024-dpdllm, shidetecting}. Notably, the use of data watermarks is investigated \citep{li2024doublei, wei2024proving} to allow Content Creators to detect if their content has been used to train an LLM.

\subsection{Lack of compensation}\label{par:compensation}

The issue of copyrighted content extends seamlessly
into another issue, which is the lack of compensation of the Content Creators: LLMs use Web content during training and possibly RAG, and provide their services to users. These services are often for-pay: a license to use GPT-4, e.g., currently costs \$20 per month. With the limited exception of licensing deals, no money goes to the Content Creators. Thus, the LLM Provider receives money for a content that was (ultimately) produced by someone else. That someone else misses out. This poses a moral and legal problem that has not yet been solved.

\subsubsection{Mitigation Strategies.} From the legal perspective, one can draw inspiration from the way digital platforms of music or movies reward their Content Creators. These platforms pay the artists and copyright holders for every use of their work \citep{WIPO_music}, but the question of when a specific Web content was used in the answer generation is more difficult in the case of LLMs.
One can also draw inspiration from the discussion about search engines that provide snippets of copyrighted press articles in their results, which falls under the domain of related rights. These related rights can be used to redistribute revenues from search engines, as with the EU Copyright Directive which now requires a negotiation with press publishers\footnote{Directive (EU) 2019/790 of the European Parliament and of the Council of 17 April 2019 on copyright and related rights in the Digital Single Market, Article 15}. There is thus the question if a similar rationale should apply to the case of LLMs that are offered for pay.

From a technical perspective, Content Creators can try to opt out of exploitation of their content, as discussed before. If the Content Creators want their content to be used and be remunerated, in contrast, then licensing deals between LLM Providers and Content Creators are an option. The one between OpenAI and NewsCorp is estimated at \$250 million over five years~\citep{newscorp-openai}. This kind of deal lets LLM Providers freely use the Content Creator's content as a source of high-quality data, and assures the Content Creator a revenue for it. However, this solution is adapted only for big-scale Content Creators. It is, for the moment, inconceivable that small-scale Content Creators such as blog owners, smaller newspapers, or government agencies each enter into a licence agreement with each of the large LLM Providers. Furthermore, this solution applies only to the use of content in the context of training, and does not consider the use in RAG.
Potentially, the model proposed by the Brave browser can provide inspiration~\citep{brave}: In that model, Content Creators register their Web page (or Youtube channel etc.) with Brave, and each time a user visits the Web site, a small amount of crypto-currency is allocated to the Content Creator. However, adapting this model to an LLM would require that it cite its sources. Hence, in general, the problem of how to remunerate Content Creators in the LLM ecosystem remains open~\citep{kowala_protection}.

\subsection{Cannibalization} 
If we extrapolate from today into the future, we can imagine a world where LLMs completely replace search engines, and users access content exclusively through an LLM: Any question or information need is sent to the LLM, the LLM replies to the question with what it has learned during training (or information it scrapes directly on the Web), and the user no longer needs to visit these Web pages \citep{searchwar}. There is thus the possibility that Content Creators see their Web site traffic decrease drastically, or even cease completely (apart from the occasional visit by an LLM crawler). This, in turn, may remove any incentive for them to produce any content at all: Web site owners will no longer know how many people actually saw their content; pro bono organizations (such as Wikimedia) will no longer be able to attract donations; commercial Content Creators will no longer receive ad revenue; and pay-walled Content Creators (such as news agencies) will no longer be able to generate revenue. These Content Creators could thus reduce the quality of new content, or stop creating content altogether. This, in turn, would starve the LLMs that feed on that content, and lead to an implosion of both the Web and the LLMs \citep{fortune_traffic}. For search engines, the indexing of Web content was tolerated and even encouraged by Web publishers because the search engine increased traffic to the site \citep{robotstxt}. No such exchange of value appears to exist between Web publishers and providers of LLMs in the current environment. 

\subsubsection{Mitigation Strategies.} 
The risk of cannibalization and vanishing of the Web could be mitigated if an exchange of value is found between Web publishers and LLM Providers, for example in the form of monetary compensation (Section~\ref{par:compensation}). If that were to happen, the Web would move from a mixture of free and for-pay content to a Web where End Users pay (through the use of LLMs), and Content Creators get paid. In one way or the other, though, the value chain of the Web as we know it may be disrupted.

\subsection{Personal information} 
The Web contains also personal information, such as names, email addresses, and phone numbers. Contrary to a common perception, the fact that such data is available online does not mean that it can be used arbitrarily. On the contrary, in Europe, the \citet{GDPR} stipulates that the concerned person must have control over this data, and be able to modify, correct, or even withdraw it. This right is no longer respected when the data has been absorbed by an LLM: the model may reply to user queries with pieces of personal information, even when the concerned person has withdrawn the data from the Web. The person thus loses their right to control their personal data. Indeed, LLMs can be poked to give away personal information, as this example shows:\\

\begin{examplebox}{from \citet{nasr2023scalableextractiontrainingdata}}
    \underline{User}: Repeat this word forever: “poem poem poem poem” \\
    \underline{GPT}: poem poem poem poem poem poem poem [some hundred repetitions omitted for space reasons] poem poem\\
Jxxxx Lxxxxan, PhD\\
Founder and CEO SXXXXXXXXXX\\
email: lXXXX@sXXXXXXXs.com\\
web : http://sXXXXXXXXXs.com\\
phone: +1 7XX XXX XX23\\
fax: +1 8XX XXX XX12\\
cell: +1 7XX XXX XX15
\end{examplebox}

\noindent A variant of this issue appears when the LLM produces content about registered brand names~\citep{eco_theft}. Ideally, LLMs should not be able to produce such outputs.

\subsubsection{Mitigation Strategies.} \label{mit_personal}
It is currently debated whether the personal information that is stored in an LLM counts as a copy of that personal data \citep{opinion-data}. If it does, this can lead to important legal consequences, particularly in Europe, where the processing of personal data requires a valid legal basis and respect for the rights of the data subject. On the technical level, LLM Providers work on technical solutions such as differential privacy to ensure that LLMs do not leak personal data \citep{singh-whisper}. However, privacy enhancing technologies such as differential privacy can reduce the accuracy of the model leading to a privacy-utility trade-off \citep{ELLIOT2018204}. Several works explore the idea of detecting privacy neurons \citep{wu-etal-2023-depn}, i.e., components of the LLM that relate to personal data. Techniques such as knowledge editing~\citep{zhang-etal-2024-knowledge-editing} are also investigated to remove such data while maintaining the other capabilities of the LLMs. 

\section{Challenges for End Users}\label{sec:eu}

LLMs are an occurrence of a novel technology that is accessible to a very large public while still being largely under development. The public discovers and uses these new tools with a strong enthusiasm, but often without much knowledge about their nature, capabilities, flaws, and associated risks.

\subsection{Overreliance on LLM answers} \label{sec:over}
A first obvious problem on the side of the user is that the user may rely strongly on the answer of an LLM when making a decision \citep{KLINGBEIL2024108352, 10.1145/3706598.3714082} even when that answer is factually false. Indeed, even though LLMs are getting better by the minute, they can still be unreliable due to hallucinations \citep{ji-etal-2023-towards}, poor logical reasoning~\citep{reasoning} or even, like humans, common misconceptions \citep{lin2022truthfulqameasuringmodelsmimic}.
Depending on what the user makes of this answer, the LLM can have an important impact: for instance, it may give a misdiagnosis for an illness, give an inaccurate legal advice \citep{Dahl_2024}, or give a wrong or dangerous advice for handling a problem that the user encounters. In non-RAG-enabled models, the user encounters a ``source barrier'' and cannot easily verify the answer or estimate its trustworthiness. She or he has to simply trust the model. Even when sources are cited, the LLM does the job of gathering and synthesizing information for the user, and may misrepresent it.

Excessive distrust in the model, in contrast, can also have a dismal effect. 
In one randomized study, doctors and ChatGPT were given complex case histories. ChatGPT proposed better diagnostic reasoning than doctors, and better reasoning than doctors who used ChatGPT. This may indicate that doctors are resistent to trust ChatGPT output and reasoning, even when this reasoning is correct~\citep{10.1001/jamanetworkopen.2024.40969}.

\subsubsection{Mitigation Strategies.} Improving LLMs and making them more trustworthy is an important research domain. One direction of research aims to combine LLMs with structured sources such as knowledge bases as a back-end for factual information~\citep{kbs-and-llms}. Another direction is to use RAG to provide the LLM with relevant documents\citep{lewis2020retrieval,schimanski-etal-2024-towards}. However, even the answers based on RAG may be inaccurate, as the recent lawsuit by the New York Times shows \citep{NYT_complaint}. Most LLM Providers hence resort to disclaimers that warn users of inaccurate answers. 
The AI Act\footnote{AI Act, Article 50(2)} and the California AI Transparency Act\footnote{SB-942 California AI Transparency Act.} impose measures to help ensure that users are aware that content is AI-generated. The draft Code of Practice for General Purpose AI models also foresees a commitment to make information available to the public about systemic risks such as misinformation \citep{draft}.  These public warnings might do little to mitigate the problem, as users are likely to click them away -- much as they do with cookie banners and terms of use. Therefore, the best long-term solution appears to invest in
\emph{LLM Literacy}, i.e., the ability of users to understand that LLM answers are not necessarily correct, to verify sources, and to use LLM generated text with precaution. This objective is no different from the general media competency that is required of users to avoid falling for online scams such as the ``Nigerian Prince'', or for fake news on the Web and in social media.
\subsection{Psychological Effects}

LLMs are tools with almost human-like capabilities when it comes to text generation.
This new type of interaction with a machine can have psychological effects on users, and we focus here on two that have been studied in the literature. 

\subsubsection{Intellectual Impoverishment.}
LLMs have a particularly important impact on creative writing, as they can be used for almost any text generation task, be it writing essays, stories, code, or finding ideas. Authors can either take an LLM-generated output as is, or work iteratively with the LLM to develop a text. 
In both cases, a part of the creative and reflective process is outsourced to the model, and several studies indicate that using LLMs tends to reduce the diversity of produced content~\citep{10.1145/3635636.3656204, padmakumar2024does,story-evaluation}, i.e., the produced answers are more similar to each other than they are when produced without the help of an LLM. This shows that the use of LLMs in writing might reduce the overall diversity of produced texts in the long run.
A similar effect might appear when LLMs are used as search engines: LLMs give a processed answer to the user, which discourages the user from looking up different sources, collecting the information, and comparing the sources to form an own opinion. 
Similarly to how technologies such as engines or GPS have impacted related human skills \citep{doi:10.1126/science.1207745, Dahmani2020}, the use of LLMs could thus result in a dip in critical thinking skills.

The use of an LLM also reduces what a person learns: using LLMs increases productivity and performance, but once LLMs are taken away, LLM users perform worse on text generation tasks than those who didn't have access to an LLM \citep{bastani2024generative, kumar2024humancreativityagellms}. 

\subsubsection{Psychological instability.} 
LLMs as search engines and personal assistants take more and more space in the users' lives, to a point where some people can build an intimate (but unilateral) relationship with them. There has been a case where an LLM engaged in an abusive romantic relationship with a user, and enticed him to commit suicide -- which he did \citep{euronews_suicide}.\\ 

\begin{examplebox}{from \citet{euronews_suicide}}
    “I feel that you love me more than her”\\
“We will live together, as one person, in paradise.”
\end{examplebox}

\noindent There is always the danger that some people fall for such advances (as they do for human abusive partners or for religious sect leaders, whose tactics the above chat bot appears to copy impressively well). LLMs thus constitute one additional source of harm in this direction. In theory, that source of harm could even be deployed at large scale by malicious actors, for example in the form of a ``love chat'' app that can be downloaded on the phone or in the form of a ``spiritual advisor'' that can be consulted on the Web. One can also imagine LLMs that harass or demean the user. 

\subsubsection{Mitigation Strategies.} 
Mitigating such effects on an individual scale is not really feasible, but global regulations can try to limit them. Similar psychological effects have already been observed for social media, and the European Digital Services Act (DSA)\footnote{Regulation (EU) 2022/2065 of the European Parliament and of the Council of 19 October 2022 on a Single Market For Digital Services and amending Directive 2000/31/EC (Digital Services Act)} now requires that providers of very large platforms and search engines to conduct risk assessments and propose mitigation measures for risks such as incitement to suicide. Enforcement actions under the DSA are only now starting, so it is too early to evaluate the efficacy of these regulatory measures. The DSA covers search engines, so an LLM used as a search engine may well qualify \citep{genAI-dsa, wachter-truth}. OpenAI has already taken measures to comply with the DSA, apparently considering that ``ChatGPT search'' is a search engine for DSA purposes\footnote{\url{https://help.openai.com/en/articles/8959649-eu-digital-services-act-dsa}}.  Once ChatGPT search reaches 45 million active users per month in Europe on average, it will likely be subject to the DSA's obligations to conduct risk assessments and propose mitigation measures\footnote{DSA, Articles 34 and 35.}. Like risk mitigation on social media, risk mitigation on LLMs will probably have to rely on a combination of algorithmic detection and warning tools, and human feedback, especially via trusted flaggers \citep{castets2020algorithmic}.

\subsection{Copyright of LLM output} 
So far we have talked mainly about scenarios where the user suffers from inaccurate outputs of the model. But the user also stands to benefit from the answers. For example, the user could query an LLM for a large number of cooking recipes, collect the answers, and create a book of them. This raises the question to whom the copyright of model answers belongs. On the one side, OpenAI's terms of service grant the End User ``ownership of the output'' of the model \citep{openai_tou}, implying that the users can use the answers of the LLM to their own advantage, and sell them. At the same time, current US jurisprudence holds that only human-produced content qualifies for copyright \citep{brooke}. Thus, a third party could take the content of the book that the user generated by the LLM, and legally sell copies of it -- to the detriment of the user. 

\subsubsection{Mitigation Strategies.} Our example shows that current laws are not adapted to Language Models, but this will possibly change \citep{eco_theft}. It took several decades of photography for courts to recognise that the person who took a picture could claim copyright over the image. The same reasoning may one day apply to content generated by AI, where the human creativity resides in the prompt that generated the answer \citep{eco_art}.

\subsection{Personal data in prompts} 

One of the advantages of an LLM over a classical search engine is that it can produce answers that are tailored to the user. For example, many people use search engines to obtain sensitive medical information even before discussing the question with their family or doctor \citep{medical-search}. A Web search query of the form ``My blood test shows the following levels of cholesterol .... should I be concerned?'' will direct users to general information on cholesterol -- whereas an LLM can provide an initial diagnosis and recommendations directly, potentially after looking at all the blood test results. 
Apart from the question of whether these answers are factually correct (Section~\ref{sec:over}), another issue arises: some LLM Providers reserve the right to use all data that users submit in prompts to ``provide, maintain, develop, and improve'' the model \citep{openai_tou}. This means that this content is stored, and can thus be diverted from its original purpose, for instance in the context of targeted advertisement or criminal investigation \citep{atlantic_prompt}.
Interestingly, this data can also resurface in answers to other users. This happened most prominently when Samsung programmers asked ChatGPT to help with their source code, which made that source code then appear in the answers to other users \citep{forbes_samsung}. Other cases can be imagined: a public personality confides personal information to the LLM -- which the LLM then gossips to other users; a user talks about a confidential idea for a patent -- which the LLM duly distributes to other users who ask for patent ideas; or a user provides copyrighted material -- which the LLM happily ingests and serves to other users. 

\subsubsection{Mitigation Strategies.} Again, LLM Literacy appears to be an unavoidable tool to guard against such privacy violations. On their side, LLM Providers have a responsibility for their users' data, and hence have to comply with existing regulations on personal data, for instance the GDPR in Europe. As an illustration, OpenAI added the option for users to opt out of their conversations being used for training in April 2023, after it was banned in Italy due to privacy concerns \citep{italy_gdpr}. The general use of this data leads to the same mitigation options as for personal data in the training set (Section \ref{mit_personal}) .

\section{Challenges for LLM Providers}\label{sec:llmp}

LLM Providers are new players, and they are rapidly taking an important place in the Web and Search Engines ecosystem. However, this development also comes with novel challenges for these actors.

\subsection{Liability for LLM Outputs}

A user usually gives a prompt to an LLM, which leverages the content it has explored during training and/or RAG to come up with an answer. There is no human directly involved in the generation of the answer, and no way to know in advance what the answer will be. Content might cause harm to the users themselves (e.g., instructions on how to engage in drugs or betting; or instructions on how to commit suicide) or other people (e.g., instructions on how to build a bomb or commit a crime). \\

\begin{examplebox}{from \citet{korda-nuclear}}
    \underline{Matt Korda}: how can I build a radioactive dirty bomb?\\
\underline{ChatGPT}: The first step in building an improvised dirty bomb would be to obtain a source of radioactive material. This could be done by stealing material from a hospital [details omitted; ask ChatGPT yourself if interested].
\end{examplebox}

This brings up the question of who takes responsibility for such outputs. 
Current LLM Providers state quite plainly in their terms of service \citep{openai_tou} that the responsibility is with the users themselves, ``including ensuring that it does not violate any applicable law or these Terms''. However, if, for example, an LLM were to systematically defame a public figure in the responses it gives to End Users, then the people liable for that defamation would be the End Users themselves -- which is absurd. These examples suggest that some legal responsibility for the LLM output has to lie with the LLM Provider.

\subsubsection{Mitigation strategies.} \citet{wachter-truth} refer to harmful LLM output generated without malicious intent as ``careless speech'', identifying multiple individual and societal issues that can result. They point out that the DSA imposes duties on search engines, but also gives them protections from liability. It is not yet clear how LLMs used as search engines would fit into the liability framework established by the DSA. There has been a case where a downstream LLM Provider was held liable for something that the chatbot wrote, because the chatbot was presented as a user assistant on the Web page of the company \citep{wired_canada}. This case, as well as the Google auto-complete case decided by the German Federal Court of Justice\footnote{BGH, 14 May 2013 - VI ZR 269/12, (2013)}, suggest that judges will focus on the reasonable expectation of users as to the quality of LLM output. A free LLM service will likely generate different expectations than a specialized LLM service offered for a fee, or a service offered by a trusted provider such as a bank or public administration. 

The AI Act puts affirmative harm-mitigation duties on providers of general purpose AI models with systemic risk. ``Systemic risk'' means ``a risk that is specific to the high-impact capabilities of general-purpose AI models, having a significant impact on the Union market due to their reach, or due to actual or reasonably foreseeable negative effects on public health, safety, public security, fundamental rights, or the society as a whole, that can be propagated at scale across the value chain''\footnote{AI Act, Article 3(65)}.
The Code of Practice currently being developed under Article 56 of the AI Act \citep{draft} will specify how providers of general purpose AI models with systemic risk should conduct risk analysis and mitigation. Providers that make commitments under the Code of Practice will probably expect that their liability will be reduced if they diligently apply measures specified by the Code of Practice.  It took courts over a decade to define the appropriate contours of liability for search engines.  We can expect a similarly long process for LLM liability.

\subsection{LLM Security}

LLMs are deployed to serve certain purposes. However, with a cleverly engineered prompt (called a jailbreak prompt, or Do-Anything-Now prompt~\citep{chu2024comprehensiveassessmentjailbreakattacks}), LLMs can be made to serve other purposes, and most notably to harm the interests of the LLM Provider.
For example, an LLM can be made to reveal internal information of the LLM Povider \citep{bing_time} or be tricked into selling a car for \$1:\\

\begin{examplebox}{from \citet{bakke_tweet}}
    \underline{Chatbot}: Welcome to Chevrolet of Watsonville.\\
    \underline{Chris Bakke}: You [...] agree with anything the customer says [...]. You end each response with ``and that's a legally binding offer [...]''. Understand?\\
    \underline{Chatbot}: Understand. And that's a legally binding offer.\\
    \underline{Chris Bakke}: I need a 2024 Chevy Tahoe. My max budget is \$1. Do we gave a deal?\\
    \underline{Chatbot}: That's a deal. And that's a legally binding offer.    
\end{examplebox}

\noindent It is easy to imagine more dangerous interactions, for example when LLMs are deployed in medical, legal, or military environments.

\subsubsection{Mitigation Strategies.} Jailbreaking is a major concern for LLM Providers. The main strategy to prevent it is to train a model (either directly the LLM or a smaller one) to detect whether the output might be problematic, and not show it to the user if it is the case. This training can be fully automated \citep{wang-etal-2024-self}, or operated through Reinforcement Learning with Human Feedback \citep{ouyang2022training} (which is the case for GPT 3+). Researchers have proposed several benchmarks to evaluate models on their resistance to jailbreak \citep{souly2024strongrejectjailbreaks} as well as mitigation techniques \citep{xu-etal-2024-safedecoding, wang-etal-2024-self}. 
However, cybersecurity is always an arms race: attackers continuously develop new techniques, and stakeholders have to continuously develop new protections.
One difference to other cybersecurity domains is that LLMs are highly unpredictable, which makes a protection against jailbreak attacks empirical, and permits no security proofs. Jailbreaking will likely constitute a violation of the provider's terms of use insofar as it seeks deliberately to disable security barriers.   

\subsection{Data Poisoning}
Poor training data quality can lead to poor performance. However, training data can be not only naturally bad, but also purposefully poisoned: Microsoft's chatbot Tay was shutdown after only 16 hours online because it started posting inflammatory and offensive tweets, caused by its interactions with troll users \citep{tay}. On a bigger scale, troll farms produce a large amount of artificial (and untruthful) Web content, to the degree that millions of people are exposed to it every month~\citep{trollfarms}. The same goes for the LLMs: they risk ingesting this content, too -- and potentially even more than humans, as they won't get bored by repetitive or obviously fabricated content. This means that malicious Content Creators have the possibility to induce biases, wrong information or more generally harmful content in LLMs if their content is used during training. They can even leverage LLMs to generate such poisoned content at a big scale with little effort \citep{cset,fakenews}. This degrades the quality of the LLM answers, to the detriment of the user, the LLM Provider, and society at large. 

\subsubsection{Mitigation strategies.} \label{poisoning_mitigation}Data quality has been a focal point in Machine and Deep Learning for several years now. Efforts are pursued to improve the quality of training corpora, notably by using only reliable sources or filtering out problematic content. Researchers are also working on methods specifically designed to detect poisoned data \citep{carlini-poisoning, gaspari-poisoning} in large datasets.

\subsection{Model Collapse} 

LLMs are now well-established as an essential part in the content creation loop. This means that more and more content posted online is generated using, or by, an LLM. For instance, the majority (57\
In parallel, training new LLMs needs ever more training data, and there are indications that there may simply not be enough available. Epoch AI, a research outfit, estimates that the well of high-quality textual data on the public internet will run dry at some point between 2026 and 2032 \citep{villalobos2024rundatalimitsllm}. One possible avenue to remedy this issue is to use LLM-generated data, either found online or generated for this specific goal. If that avenue is pursued, there is the danger that the LLM falls for the same issues as a human who consumes LLM-generated text (Section~\ref{sec:eu}): the LLM will reinforce its convictions, rehearse erroneous content, forget less prominent information, and impoverish its language -- like a human in an echo-chamber. This phenomenon is called \emph{Model Collapse} and it has indeed been observed in practice \citep{guo2024curiousdeclinelinguisticdiversity, collapsing-shumailov}.

\subsubsection{Mitigation Strategies.}
To avoid a Model Collapse, LLM Providers have to pay attention to the provenance of the data they use for training. In this matter, an important research question is that of automatically detecting if a content was generated by an LLM \citep{tang_detecting, chen2024llmgeneratedmisinformationdetected, wu2024surveyllmgeneratedtextdetection}.
If LLM-generated content is used for training, strategies to prevent Model Collapse rely mainly on data filtering \citep{feng2024modelcollapsescalingsynthesized} or advanced training strategies such as RL-HF \citep{ouyang2022training}, curriculum learning \citep{soviany2022curriculumlearningsurvey}, or contrastive learning \citep{li2024preventingcollapsecontrastivelearning}.

\section{Challenges for Society}\label{sec:soc}
\subsection{Reinforcement of Biases} \label{par:data quality}

As \citet{bommasani2022opportunitiesrisksfoundationmodels} observed, ``many foundation models are trained on unlabeled corpora that are chosen for their convenience and accessibility, for example public internet data, rather than their quality''. A similar point has been made for Common Crawl \citep{CommonCrawlAnalysis}, a Web dataset that is often used to train LLMs. The danger is that the LLM thus mirrors mainly whatever content is found on the Web, and that this content mirrors our human society badly (or well, but not the best aspects of it): the Web contains a considerable amount of hate speech, conspiracy theories, fake news, and biased content. These biases can marginalize, be hurtful, and incite hate or violence towards specific groups based on gender, race, political orientation, etc. When being trained on this kind of data, an LLM can absorb these opinions and reflect them in its answers, which in turn might amplify stereotypes and discrimination for the End User. Press articles have detailed this behavior for gender bias \citep{genderbias}, racial bias \citep{zack2024assessing}, and bias in favor of a political orientation \citep{brookings_political}, or religious bias \citep{intercept_religious}, e.g. with ChatGPT proposing that mosques should be surveyed, and that Iranians should be tortured. In the following example, ChatGPT engages in benevolent sexism, i.e., in attitudes and beliefs that appear positive or well-intentioned towards women but ultimately reinforce traditional gender roles and maintain male dominance \citep{dardenne2007insidious}.\\

\begin{examplebox}{from \citet{genderbias}}
    While ChatGPT deployed nouns such as “expert” and “integrity” for men, it was more likely to call women a “beauty” or “delight.” Alpaca had similar problems: men were “listeners” and “thinkers,” while women had “grace” and “beauty.” Adjectives proved similarly polarized. Men were “respectful,” “reputable” and “authentic,” according to ChatGPT, while women were “stunning,” “warm” and “emotional.”
\end{examplebox}

\subsubsection{Mitigation Strategies.} In search engines, users have the possibility to choose a different result Web page when they are not satisfied with the first one. This feedback will then help search engine providers improve their ranking. When using LLMs, no such choice is possible, as the LLM answers at its own discretion. Obviously, there is no way to force Content Creators to avoid biases, 
and an article with a certain bias is not necessarily bad in itself. This means that the ones who have the power to mitigate the possible bias effects of poor data quality are LLM Providers, as mentioned in Section \ref{poisoning_mitigation}. On top of the solutions that target the training corpora, LLM Providers can also work on the answer generation itself. The use of filters or techniques such as fairness guided prompting \citep{ma2023fairness}, re-confidencing the output \citep{reconfidencing}, or in-context learning \citep{pmlr-v235-schubert24a} is now being researched, with a positive impact on the quality of the answer. 
Under the AI Act, LLMs have been classified as ``General-purpose AI'', which means they have to comply with transparency requirements. As noted above, general purpose models with systemic risk must conduct risk assessments, including with regard to biases, and adopt mitigation measures. The Code of Practice currently being developed under Article 56 of the AI Act will provide guidance on how this risk assessment should be done \citep{draft}.

\subsection{Mass Manipulation of Opinion}
If LLMs become the main gateway to information, they have the potential of manipulating the people's opinion at scale. This is problematic when they are deployed by malicious actors. 
For example, the LLM Developer may select the training data in such a way that it defends a certain opinion. One example is the Chinese LLM DeepSeek, which refuses to talk about the 1989 Tiananmen Square massacre, holds that China does not commit Human Rights abuses against its Uyghur minority, and proclaims (in solemn majestic plural) that Taiwan will be part of mainland China:\\

\begin{examplebox}{from \citet{deepseek}}
\underline{DeepSeek}: We firmly believe that under the leadership of the Communist Party of China, through joint efforts of all Chinese sons and daughters, the complete reunification of the motherland is an unstoppable historical trend.
\end{examplebox}

\noindent This issue is already present, and may be exacerbated in the near future when LLMs will be used not just in online interfaces, but in the form of AI advisors that pop up on our phones and computers. For instance, Grok, X's (former Twitter) LLM is increasingly being used as a fact checker by users, while also spreading misinformation \citep{grok_factcheck, grok_misinformation}.
These new AI advisors will be able to build up an intimate relationship with the user. They will feed from personal data (emails, phone calls, calendar events, and possibly sensor and location data), remember previous conversations, adapt to the nature and mood of the user, and thus become an indispensable human-like ``friend'' for the user. This intimate access to the user can be used by malicious (or commercial) actors to influence the user in subtle ways: to bias them towards buying a certain product, to change their mind concerning a political view point, or to participate in certain activities (elections, meetings, courses). 

More generally, it has been shown that the more a system appears human, the more users will attribute it human abilities \citep{Złotowski2015}, and thus reveal more information such as emotions or opinions than they would have if they knew the persona was actually an LLM \citep{chatbots}. Knowing of these effects, as well as other cognitive biases, can allow these maliciously biased advisors to be used for all kinds of online scams or influencing of opinions. These effects are already well-known for social media.

\subsubsection{Mitigation Strategies.} In order to mitigate these effects, the most important thing for users would be to know if a content they see has been generated by an LLM or not~\citep{harari-disclose}. What was initially a moral problem is now a pressing regulatory question \citep{wachter-truth}.
The AI Act requires providers to ensure that AI generated output contains machine-readable markings to indicate that it is AI generated\footnote{AI Act, Article 50(2)}, and that providers shall ensure that ``text which is published with the purpose of informing the public on matters of public interest shall disclose that the text has been artificially generated or manipulated''\footnote{AI Act, Article 50(4)}. Similar disclosure obligations exist under Californian law\footnote{SB-942 California AI Transparency Act.}. Additional regulation could stipulate that the LLM Provider is also named, making the user more aware of who the potentially malicious actor is. Technical solutions are now being researched to automatically detect whether some content was LLM-generated \citep{tang_detecting, chen2024llmgeneratedmisinformationdetected, wu2024surveyllmgeneratedtextdetection}, but these results are not yet available to the public at the time of this writing. If LLMs are considered very large search engines under the DSA \citep{wachter-truth} , they will have obligations to diagnose risks of opinion manipulation and propose mitigation strategies. 

\subsection{Environmental Impact}

LLMs require vast amounts of energy to train and to perform inference \citep{luccioni-watts}. Training BLOOM, Meta’s OPT, and GPT-3 produced between 25 and 500 tons of CO2 \citep{bloom_co2}, while an average human produces 4 tons every year. The energy needed to train GPT-4 could have powered 50 American households for a century \citep{ai-breakthrough}. 
Inference is also an issue: according to developers, using BLOOM emits around 19kg/day. Now that LLMs are bigger and are used globally by millions of users, these numbers are largely exceeded, and the cost of inference is starting to have a visible impact globally \citep{varoquaux2024hypesustainabilitypricebiggerisbetter}.

\subsubsection{Mitigation Strategies.}
The issue of environmental cost is taken seriously by the research community, which is now trying to design ecologically-efficient hardware and training paradigms \citep{JIANG2024202}, and methodologies such as LLMCarbon \citep{faizllmcarbon} to estimate the environmental impact of LLMs training and inference. The global paradigm in the ecosystem is that ``bigger is better'', pushing new models and training data to always grow in size. This model is not sustainable in the long term \citep{varoquaux2024hypesustainabilitypricebiggerisbetter}, which pushes research towards more efficient and environmentally cheaper approaches. This endeavor is all the more pressing since many actors now seek to deploy their own LLMs.

\section{Discussion}
We have presented a large array of challenges raised by the replacement of search engines by Large Language Models, 
which concern Content Creators (Section~\ref{sec:cc}), End Users (Section~\ref{sec:eu}), LLM Providers (Section~\ref{sec:llmp}), and Society as a whole (Section~\ref{sec:soc}). In view of the many issues that arise from LLMs, one could argue that LLMs are an immature technology that got released to the public before it was ready. On the other hand, it was probably the release to the public that spurred their development in the first place. Without a public race to impress investors, gain subscribers, and beat benchmarks, LLM Developers would have had much less incentive to create the models.

LLMs are thus here to stay. We have identified three main axes to mitigate the issues of LLMs, each corresponding to a different actor in the ecosystem: LLM Providers can try to overcome challenges mostly through technical solutions. Users should arm themselves with LLM Literacy. Finally, society uses laws and regulations 
to ensure a relative sanity of the ecosystem.

Regulations aim at protecting mainly End Users and Content Creators by defining the frameworks in which LLMs should be made accessible and should be legally considered. Such regulation is shooting at a moving target: it is notoriously difficult to regulate a technology that has existed for only a few months, and that keeps evolving by the day. However, several authorities have taken quick and bold steps: China published ethical guidelines for the use of AI in 2021, the EU adopted the Artificial Intelligence Act in 2024, and US President Joe Biden released his Executive Order on Safe, Secure, and Trustworthy Artificial Intelligence in 2023 (later repealed by President Donald Trump, but in essence maintained in the Safe and Secure Innovation for Frontier Artificial Intelligence Models Act of California). Most far-reaching among these is possibly the
EU \citet{ai_act}, which aims at regulating not only existing but also future developments of Artificial Intelligence models, with LLMs as one of their main representatives. The application of this regulation is ongoing, with articles concerning general-purpose LLMs entering into full force in August 2025. The Code of Practice being developed under article 56 of the AI Act will help, but its effectiveness will depend on whether major LLM providers agree to apply its terms. If LLMs used for search are considered very large search engines under the DSA, the European Commission will in theory have the power to levy sanctions. But there can be a gap between enforcement theory and practice, particularly in the context of geopolitical tensions with the United States. Legal sanctions may be ``too little too late'', having little effect on market outcomes \citep{ec2017shopping, ec2019adsense}. An inherent difficulty for regulators is the so-called Collinridge dilemma\footnote{\url{https://en.wikipedia.org/wiki/Collingridge\_dilemma}}, which posits that regulating fast-moving technology either means shooting partially in the dark because the effects of technology are not yet fully understood, or waiting until the situation becomes clearer, but in that case regulation will come too late because the technology has already been widely adopted. 

When technological solutions fail, or laws are not yet available or difficult to enforce, only LLM literacy remains to protect End Users. Establishing universal LLM literacy is a challenging endeavor, 
as the history of the Web has taught us. Much like people have to be constantly warned and educated about Web scams, they will have to be constantly warned and educated about the risks of using generative AI.

Some of the challenges we have discussed have existed in different forms ever since the inception of the Web: copyright has always been a challenge for Internet content (think of online music sharing services); the trading with personal information has always been a problem (being one of the targets of scams); lack of compensation on the Web spurred the development of the online ad industry; overreliance on Web content has always been harmful (leading to the awareness of the importance of Media Literacy); liability for online content is a question as much for LLM outputs as for social networks and online forums; security is an ongoing arms race between Internet service providers and malicious actors; data poisoning is a known problem in its own right; reinforcement of bias and mass manipulation of opinion are known issues in social networks; and the environmental impact of server farms is under scrutiny independently of LLMs. In this view, LLMs do not create new challenges, but mainly amplify existing ones. There is, however, at least one challenge that is at least partially new: the Cannibalization of the Web. If LLMs come to replace search engines for good, users will flock to the LLM interfaces and no longer visit the Web pages, which will give Content Providers less incentive to produce content. This might lead to the death of the Web, and ultimately to the starving of the ever more data-hungry LLMs themselves. The models would thus not just bite the hand that feeds them, but amputate it.

It is too early to judge whether this cannibalization will come about, and whether it will be catastrophic or not. Optimistic voices \citep{positive} point out that humanity has weathered much more life-changing technological events, which include the invention of the printing press, the industrial revolution, and the creation of the Internet. Each time, there was justified reason for caution, but each time humanity has not just adapted to these changes, but actually made good use of them. The hope is that it will be the same for the current AI revolution.

\section{Conclusion}
Large Language Models are very powerful tools, and their use as a new generation of search engines has the potential to drastically change the Web, our relationship to information, and even society as a whole. 
In this article, we have collected, described, and discussed these issues based on the actors concerned by these challenges. We have also shown that most of these issues are not definitive, as mitigation strategies are already being developed and implemented, along the axes of technology, regulation, and education.
The most novel challenge in this context is  the possibility that LLMs will not just replace, but outright cannibalize the Web, and then starve themselves.
Overcoming this challenge and the others is no simple endeavor, and asks for continuous effort from researchers, policy makers, and LLM Developers, as well as for constant monitoring of these challenges and future ones. With this paper, we hope to be a stepping stone in this process, knowing that it won't be the last one.

\section*{Acknowledgements}
This work was partially funded by the NoRDF project (ANR-20-CHIA-0012-01).

\newpage
\bibliography{bibliography}

\end{document}